\documentclass[10pt,a4paper]{article}

\usepackage[a4paper,margin=2.5cm]{geometry}

\usepackage{graphicx}
\usepackage{amsmath}
\usepackage{cite}
\usepackage{setspace}
\usepackage{titlesec}
\usepackage{caption}
\usepackage{xurl}
\usepackage[hidelinks]{hyperref}

\usepackage{subcaption}

\titleformat{\section}
  {\normalfont\fontsize{11}{12}\bfseries}
  {\thesection.}
  {0.5em}
  {}

\titleformat{\subsection}
  {\normalfont\fontsize{10}{12}\bfseries}
  {\thesubsection}
  {0.5em}
  {}
  
\begin{document}

\vspace*{-1.5cm}

\begin{center}
{\fontsize{14}{16}\selectfont\bfseries
AI-based Interference Mitigation for Power-domain Spectrum Sharing among LEO Satellites
}

\vspace{0.7cm}

{\fontsize{11}{13}\selectfont
Bowen Zhang\\
University of Surrey, Guildford, Surrey, UK, GU2 7XH, bz0004@surrey.ac.uk\\
Barry Evans\\
University of Surrey, Guildford, Surrey, UK, GU2 7XH,
b.evans@surrey.ac.uk\\
Pei Xiao\\
University of Surrey, Guildford, Surrey, UK, GU2 7XH,
p.xiao@surrey.ac.uk\\
}
\end{center}

\vspace{0.3cm}

\section{Abstract}
Low-earth orbits (LEO) satellites have recently gained wide attention for their potential to provide high-data-rate services in desert, sea, and rural areas. Currently such systems operate in the Ku band on the down-link. However, inter-operator LEO satellites produce frequent In-Line-Events (ILE) creating co-channel interference between their waveforms. This situation will be exacerbated with the launch of more constellations being launched. To address this challenge, we propose a novel interference mitigation system based on power control. Specifically, satellites from one operator will adjust its transmit power to create a sufficient received-power gap for co-frequency beams, enabling successive interference cancellation (SIC) or AI-assisted algorithms on the user-side for interference mitigation, where AI requires a smaller power gap. We use the waveforms from Starlink and EutelsatOneweb to evaluate the designed system. Results show the designed system can significantly mitigate interference and increase the overall spectrum efficiency.

\section{Introduction}
Low-earth orbits (LEO) satellites have gained wide attention for their potential to provide high-data-rate services in desert, sea, and rural areas. In 2025, there are about 25,000 LEO satellites \cite{esa2025spaceenvironment} and the expected number of satellites will reach 60,000 by 2030 \cite{gao2022satelliteconstellations}. Currently such systems operate in the same frequency bands e.g. Ku band on the down-link. However inter-operator LEO satellites produce instances of In-Line-Events(ILE) creating co-frequency interference between broadband LEO operators, while interference occurs for devices with omnidirectional antenna regardless of ILE. This situation will be exacerbated with the launch of more constellations and the satellite deployment expanded internationally. As the Ku band spectrum becomes increasingly congested, to enhance throughput capacity, frequency-reuse shall be considered among different satellite operators but frequent interference becomes a constraining feature in use of the spectrum. 

To address this challenge, we aim to apply non-orthogonal multiple access (NOMA) techniques \cite{liu2017nonorthogonal} for interference mitigation when the same time and frequency resource is occupied by more than one LEO satellite company. However, because some NOMA techniques, such as multibeam
satellite architectures with three/four/seven-cell frequency reuse \cite{rao2003parametric} and dual
orthogonal polarization techniques \cite{kelmendi2026cross}, have already been deployed for intra-system capacity enhancement, while power-domain NOMA (PD-NOMA) \cite{saito2013non, chen2017optimization} remains largely theoretical \cite{zhang2024relay, zhang2023unified, wang2022noma}, this work investigates its application to inter-system interference mitigation, thereby preserving intra-system capacity. 

For interference mitigation with PD-NOMA, satellites from two operators will adjust transmit power to create a sufficient received-power gap for co-frequency beams, enabling mitigation algorithms on the user-side for interference mitigation. Due to the mobility and orbit difference of different LEO systems, the relative spatial locations of co-frequency beams vary over time, thereby changing the required power gap. Unlike prior PD-NOMA works using successive
interference cancellation (SIC), we design an AI-assisted interference mitigation network to reduce the power gap requirement and enhance the aggregate capacity of coexisting LEO systems, following the success of AI networks in channel estimation \cite{gao2023deep} and wireless sensing \cite{zhang2023csi}. To evaluate the performance of the designed system, we consider the frequency reuse between Starlink and EutelsatOneweb and use their waveform structure in the public literature, which are Single-Carrier Time-Division Multiplexing (SC-TDM) and Orthogonal Frequency-Division Multiplexing (OFDM), respectively \cite{humphreys2023signal, komodromos2026signal}. 

Our main findings can be summarized as follows:
\begin{itemize}
    \item PD-NOMA is effective for interference mitigation between frequency-reuse LEO satellite operators and can improve channel capacity after careful power control,
    \item AI-assisted interference mitigation algorithms can outperform SIC when the received-power gap meet requirements,
    \item  In the 2-UE case, the data rate of the strong signal can perform better than single-user access under $ \text{log}_{2}(1+\frac{S}{N+I})$, while the weak signal's capacity can reach $ \text{log}_{2}(1+\frac{S}{N})$.
\end{itemize}

The rest of this paper is organized as follows:
The third section will give the overview of the interference scenarios between two LEO operators and provide the mathematical formulation of received waveform. The forth section will introduce the designed AI-assisted mitigation algorithm. The fifth Section will show the experimental results.

 

\section{System Overview:}
\begin{figure}[htbp]
    \centering
    \begin{subfigure}[t]{0.3\columnwidth}
        \centering
        \includegraphics[width=\linewidth]{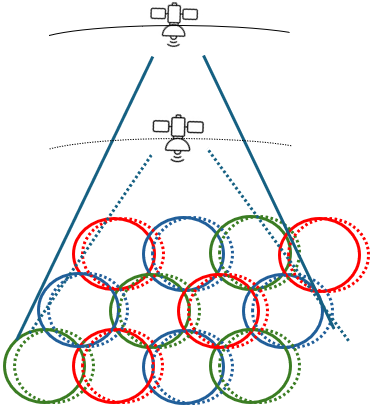}
        \caption{Interference scenario when the separation between co-frequency beam centers is small.}
        \label{fig:first}
    \end{subfigure}
     \hspace{0.02\columnwidth}
    \begin{subfigure}[t]{0.3\columnwidth}
        \centering
        \includegraphics[width=\linewidth]{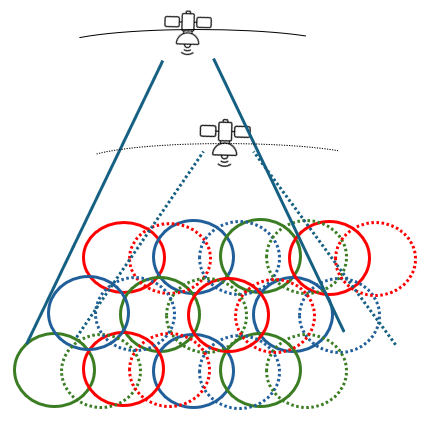}
        \caption{Interference scenario when the separation between co-frequency beam centers is large. }
        \label{fig:second}
    \end{subfigure}
    \caption{Interference scenarios for two LEO satellite systems with frequency
re-use factor 3 (FR3).}
    \label{fig:interference_overview}
\end{figure}

\begin{figure}[htbp]
    \centering
    \includegraphics[width=0.7\linewidth]{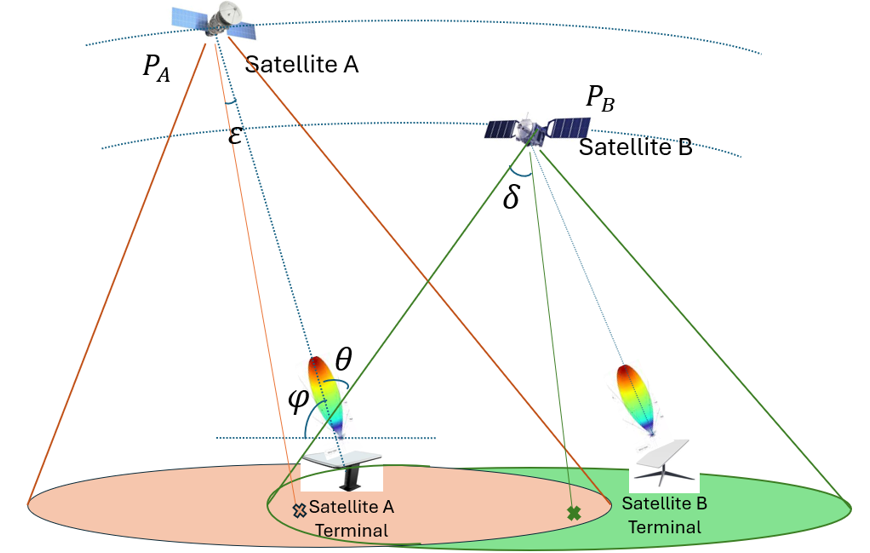}
    \caption{The illustration of the signals received at terminals in co-frequency beams from two LEO satellite operators.}
    \label{fig:signal_overview}
\end{figure}

\textbf{Interference scenarios:}
When two LEO satellite operators share frequency, their co-frequency beams will cause interference to one another. We show the interference scenarios for two LEO satellite systems with the same beam size and frequency re-use factor (e.g. FR3) in Fig. \ref{fig:interference_overview}. In Fig. \ref{fig:interference_overview}, beams with the same color share the same frequency bands while beams with solid and dashed lines belong to different operators. Co-frequency beams from the same company will not cause inter-beam interference because of their large spatial distance. However, during ILEs, interference occurs because different operators' co-frequency beams become spatially close and their signals fall within the main lobe of a terminal antenna. In addition, Fig. \ref{fig:interference_overview} shows two different interference scenarios. When inter-operator co-frequency beams are well aligned, the strengths of desired-signal and interference-signal are comparable. However, when they partially overlap or remain spatially separated, the strength of interference-signal will be lower than that of desired-signal. Scenario-dependent signal-strength variations increases the satellite-side power demand to maintain the received-power gap needed for interference mitigation.

\textbf{Received-power gap:} We then show the signals received at terminals in co-frequency beams from two LEO satellite operators in Fig. \ref{fig:signal_overview}. We will calculate the received-power between the desired signal and interference signal at terminal A. Suppose satellite A and satellite B are from two operators and their co-frequency beams are spatially close. Suppose satellite A is providing communication services to satellite A terminal with a transmit power $P_{A}$. Simultaneously, terminal A receives co-frequency interference from satellite B’s transmission intended for terminal B. Suppose Gaussian antenna is applied to both sides. Define the orbit height of satellite A and B as $H_{1}$ and $H_{2}$, respectively. Define the earth radius as $R$, the elevation angle of satellite A as $\varphi$, the angular separation between satellite B and satellite A seen from ternimal A as $\theta$, the relative angular of terminal A to the beam center of satellite A and B as $\epsilon$ and $\delta$, respectively. The power-gap (in dB) between desired signal and interference signal received at terminal A can be approximately calculated as $G =P_{A}-P_{B}+20 \text{log}⁡_{10}(\sqrt{(R^2 \sin⁡(\phi+\theta)^2+H_{2} (H_{2}+2R) )}+R\sin(\phi+\theta))-20\text{log}⁡_{10}(\sqrt{R^2 \sin⁡(\phi)^2+H_{1} (H_{1}+2R))}+R \sin⁡(\phi))+12(\delta/HPBW_{SatB} )^2+12(\theta/HPBW_{terminal A})^2-12(\epsilon/HPBW_{SatA})^2$, where HPBW denotes half-power beamwidth.

\textbf{Received waveform:} For simplicity, let the received waveform at terminal A be $\mathbf{Z}=\mathbf{X}+\mathbf{Y}+\mathbf{N}, \mathbf{N} \sim \mathcal{CN}(0, kTB)$, where $\mathbf{X}$, $\mathbf{Y}$, and  $\mathbf{N}$ are desired signal, interference signal, and thermal noise with powers $P^{A}_{A}$, $P^{B}_{A}$, and $P^{N}=kTB$, respectively; $k$ is Boltzmann’s constant, $T$ is the noise temperature, $B$ is the effective noise bandwidth, and $P^{A}_{A}$, $P^{B}_{A}$ are the received link power from satelite A and B to terminal A, respectively. The signal-to-noise ratios (SNRs) of the desired signal and interference signal are $P^{A}_{A}/P^{N}$ and $P^{B}_{A}/P^{N}$, respectively. The signal to interference and noise ratio (SINR) is $P^{A}_{A}/(P^{B}_{A}+P^{N})$.  Also, the power gap $G = 10\text{log}_{10}(P^{A}_{A}/P^{B}_{A})$. Here, we assume that two terminals are close and Doppler pre-compensation is applied at the Satellite side, therefore, the Doppler effect is ignored. We also assume additive white Gaussian noise (AWGN) channel.


\section{AI-assisted Mitigation Algorithms:}
\begin{figure}[htbp]
    \centering
    \includegraphics[width=0.9\linewidth]{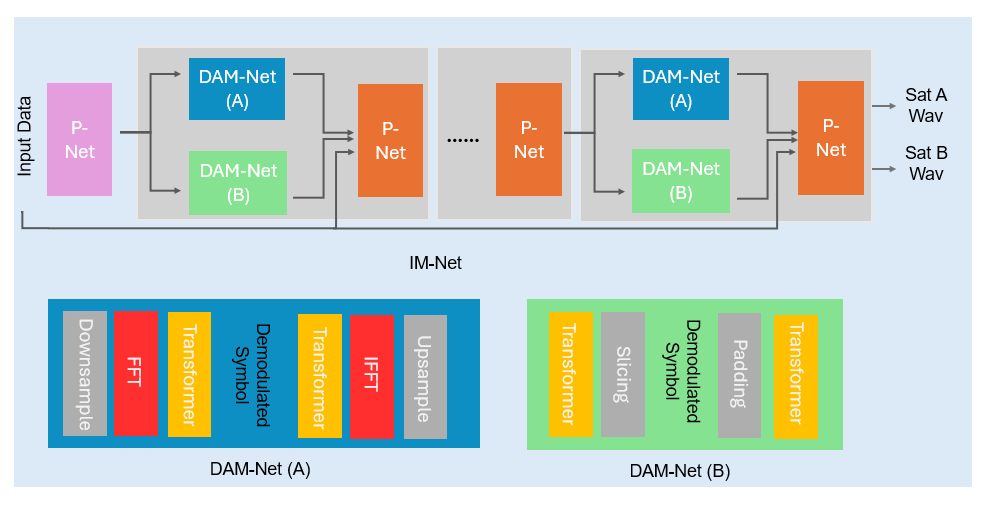}
    \caption{The overall network architecture of the interference mitigation network.}
    \label{fig:signal_overview}
\end{figure}

The SIC algorithm has been widely used in PD-NOMA \cite{saito2013non}. The idea is to process the mixture of the desired signal and interference signals iteratively, with each waveform demodulated, modulated, and removed from the mixture in the descending order of its relative signal strength. Once all interference waveforms have been obtained, they are removed from the original waveform before the final demodulation of the desired signal.

However, the performance of the SIC algorithm is restricted by the independent processing of each waveform. To enable joint demodulation of the mixture signal and improve demodulation performance, we design an AI-assisted interference mitigation network. Suppose two satellites use SC-TDM and OFDM, respectively.  Fig \ref{fig:signal_overview} shows the overall network architecture of the interference mitigation network (IM-Net). IM-Net takes the received waveform, its OFDM-structured frequency-domain representation, and the SNRs of both links as inputs. It processes them through an initial processing network (P-Net), followed by several cascaded blocks, each comprising two types of demodulation-and-modulation networks (DAM-Nets) and one P-Net, to recover the waveforms of both links. 

\textbf{Initial P-Net:} The initial P-Net is an eight-layer Transformer with 2 outputs, with each output representing the time-domain waveform of each link. The functionality of P-Net is to initially separate two waveforms in time-domain and prepare them for the post processing. 

\textbf{Processing blocks:} Each block consists of two types of DAM-Nets (i.e. DAM-Net (A) and DAM-Net (B)) and one P-Net. DAM-Net (A) is designed for OFDM waveforms. Its demodulation part consists of one resampling block to compensate for the difference of symbol rates and sampling rates between desired and interference waveforms, one FFT block to transform the waveform from time-domain to frequency-domain, and one Transformer with four layers for symbol detection. Its modulation part is designed in a reverse order, with a Transformer for modulation, an IFFT module to transform the waveform back to time-domain, and a resampling block to adjust the rates. DAM-Net (B) is designed for SC-TDM waveforms. Its demodulation part is composed of one Transformer for symbol detection at sampling rate and a slicing block to obtain symbols, while its modulation part has an asymmetric structure. Finally, the reconstructed waveforms and the input data are concatenated and fed into one P-Net to obtain the time-domain waveform for both links. The functionality of the P-Net is to remove the reconstructed SC-TDM waveform from received waveform to get interference-free OFDM waveform and remove OFDM waveform from received waveform to get inteference-free SC-TDM waveform for next processing block, respectively.  We set the number of processing blocks as three. 

\textbf{Training Process:} We use the waveforms without mutual interference, i.e. $\mathbf{X} + \mathbf{N}$ and  $\mathbf{Y} + \mathbf{N}$, and detected symbols from $\mathbf{X}$ and $\mathbf{Y}$ as training labels. They are used to train the time-domain waveforms output from P-Net and the demodulated symbols in each processing block.

\section{Experiments:}
\begin{figure}[htbp]
    \centering
    \includegraphics[width=0.75\linewidth]{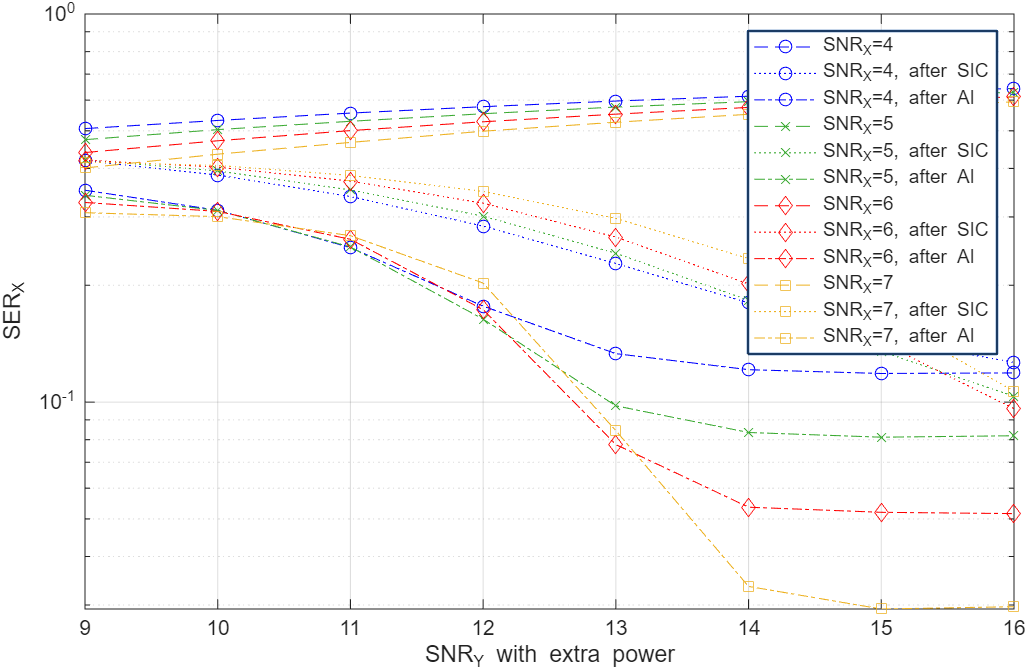}
    \caption{The SER before and after mitigation algorithms.}
    \label{fig:A_SER}
\end{figure}

\begin{figure}[htbp]
    \centering
    \includegraphics[width=0.8\linewidth]{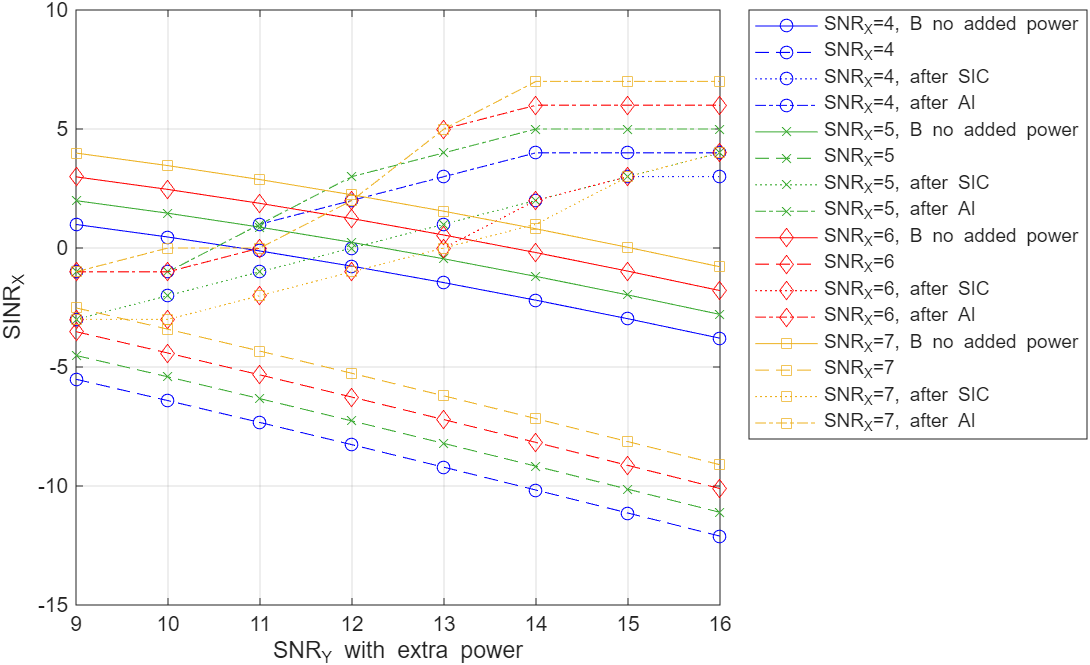}
    \caption{The SINR changes of terminal A before and after mitigation algorithms.}
    \label{fig:A_SINR}
\end{figure}

\begin{figure}[htbp]
    \centering
    \includegraphics[width=0.8\linewidth]{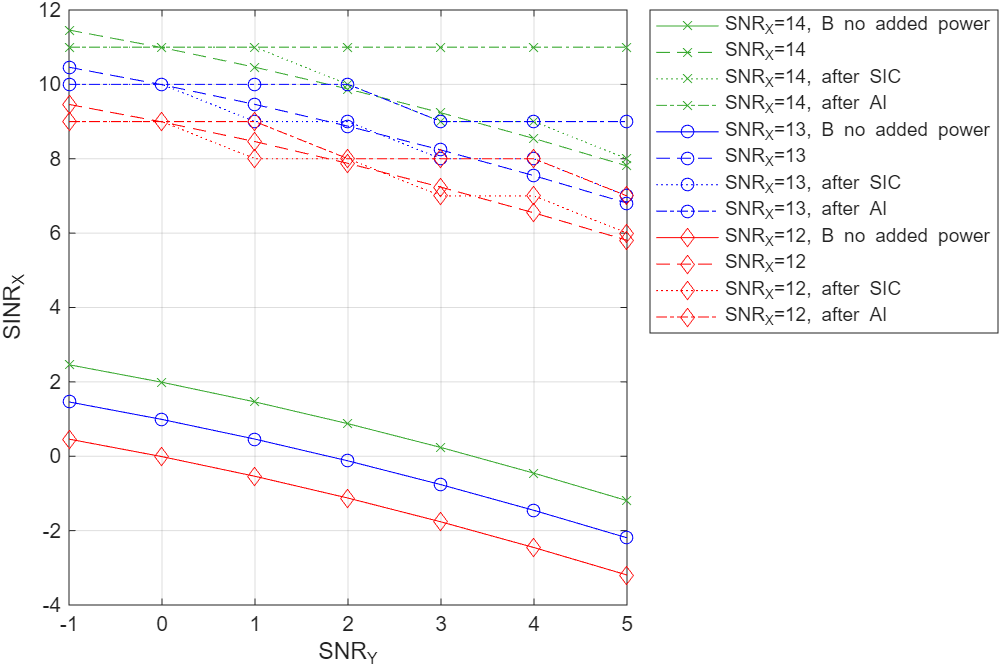}
    \caption{The SINR changes of terminal B before and after mitigation algorithms.}
    \label{fig:B_SINR}
\end{figure}
From \cite{zhao2025measuring, oneweb2016gen1}, OneWeb uses quadrature phase-shift keying (QPSK) for downlinks and the SNR of its link can range from 4 dB to 7 dB, while the restriction on the spectral power flux density from FCC for different LEO operators is the same. Therefore, $10\text{log}_{10}(P^A_{A}/P^N) \simeq 10\text{log}_{10}(P^B_{B}/P^N) \simeq 4 \sim 7 \text{dB}$ in reality. However, PD-NOMA requires a received-power gap for mitigation algorithms. Therefore, we increase $P_B$ by 9 dB. Also, as shown in Fig. \ref{fig:interference_overview}, the interference power, $P^B_{A}$ and $P^A_{B}$, gradually decreases when the distance of the co-frequency beam centers increases, we consider these changes for both terminal A and terminal B and set Satellite A and B to use SC-TDM and OFDM, respectively. 

Fig \ref{fig:A_SER} shows the symbol-error-rate (SER) of the desired signal at terminal A, $SER_X$, before and after mitigation algorithms when the SNR of interference link, $\text{SNR}_{Y}$, changes from 16 dB to 9 dB, and the desired signal power, $\text{SNR}_{X}$, changes from 4 dB to 7 dB. From Fig \ref{fig:A_SER}, $\text{SER}_{X}$ increases as the growth of $\text{SNR}_{Y}$ without mitigation algorithms but decreases after SIC and AI are utilized. The AI algorithm outperforms SIC in all conditions and can realize the complete recovery of desired waveforms when $\text{SNR}_{Y}$ is above 14 dB.

Fig \ref{fig:A_SINR} shows the SINR changes of desired signal at terminal A, $\text{SINR}_{X}$, before and after mitigation algorithms under the same condition as Fig \ref{fig:A_SER}, where 'No added power' denotes the situation when $P_B$ does not increase, i.e., $\text{SNR}_{Y} \in [0, 7]$ dB, while the 9 dB increase is considered in other lines. SINR is estimated by comparing the SERs of QPSK modulated SC-TDM waveforms in AWGN channel. As shown in Fig \ref{fig:A_SINR}, increasing $P_B$ decreases $SINR_{X}$ if no mitigation algorithm is applied. However, both SIC and AI can help increase $SINR_{X}$, and AI outperforms SIC in all cases. When $SNR_{Y}$ is above $13$ dB, $SINR_{X}$ after SIC is above the $SINR_{X}$ without power change, while $SINR_{X}$ after AI can exceed $SINR_{X}$ without power change if $SNR_{Y}$ is above $11$ or $12$ dB. Note that from \cite{komodromos2026signal}, the OneWeb satellite’s beam coverage corresponds approximately to the antenna’s −3 dB beamwidth. If we apply the same beam coverage settings to satellite A and B and consider that terminal A uses omnidirectional antenna, as long as terminal A is inside the co-frequency beam of satellite B, increasing $P_B$ by 9 dB can improve $\mathrm{SINR}_X$ at terminal A when combined with AI-based interference mitigation algorithms as $SNR_{Y}$ will be above 11 dB. However, when terminal A is outside satellite B's beam or has a highly directional antenna,  a higher yet still insufficient value of $P_B$ may adversely affect terminal A as the requirement of the received-power gap cannot be met. To address this issue, $P_{B}$ can be either further increased or decreased back to the normal value, i.e. without the 9 dB gain, when satellite B's co-frequency beams move away from terminal A. 

Fig \ref{fig:B_SINR} shows the SINR changes of desired signal, $\text{SINR}_{X}$, at terminal B when the SNR of desired signal, $\text{SNR}_{X}$, ranges from $12 \sim 14$ dB and the SNR of interference signal, $\text{SNR}_{Y}$, changes from $5$ to $-1$ dB. The SINR is estimated by comparing the SERs of QPSK-modulated OFDM waveforms in AWGN channel. From Fig \ref{fig:B_SINR}, increasing $P_B$ substantially improves $\text{SNR}_{X}$; therefore, increasing $P_{B}$ benefits both terminals. Besides, $\text{SINR}_{X}$ after SIC performs similar with $\text{SINR}_{X}$ without algorithms, while $\text{SINR}_{X}$ after AI performs significantly better. Also, AI is worth considering when $\text{SNR}_{Y}$ is above $0$ dB. However, despite the SINR improvement of waveforms, the spectrum efficiency of terminal B is still capped at 2 due to the QPSK modulation.  

To sum up, increasing the transmit power of one operator accompanied by the AI-based mitigation algorithm at the user side can effectively maintain the link quality of terminals from both operators. During this process, real-time cooperation is not required but the sharing of beam coverage plan is needed. 

\bibliographystyle{IEEEtran}
\bibliography{sample}





\end{document}